\documentclass[showpacs,floatfix,preprint,nofootinbib]{revtex4}
\usepackage{amsmath}

\begin{document}

\title{Neutrino statistics and non-standard commutation relations}
\author{A.~Yu.~Ignatiev}
\affiliation{ 
    \em School of Physics, Research Centre for High Energy Physics,
     University of Melbourne,   Australia}
\email{a.ignatiev@physics.unimelb.edu.au}
\author{V.~A.~Kuzmin}
\affiliation{ 
    ~{\em Institute for Nuclear Research of the Russian Academy of Sciences}\\
    {\em Moscow, Russia.}}
\email{kuzmin@ms2.inr.ac.ru}


\def\be{\begin{equation}}
\def\ee{\end{equation}}
\def\bea{\begin{eqnarray}}
\def\eea{\end{eqnarray}}
\newcommand{\nn}{\nonumber \\}

\begin{abstract}
Recently it was suggested that the neutrino may violate the Pauli 
exclusion Principle (PEP). This renews interest in the systematic search 
for bilinear commutation relations that could describe deviations from 
PEP. In the context of this search we prove a no-go theorem which forbids  
a   finite occupancy limit for an arbitrary system with a bilinear 
commutation relation. In other words, either the upper limit on the 
occupancy number is 1 (the ordinary fermionic case) or there is no upper 
limit at all. Some examples of the latter class include the usual Bose 
statistics, as well as non-standard quon statistics and infinite 
statistics.    
\end{abstract} \maketitle 

The Pauli exclusion principle (PEP), one of the cornerstones of physics 
and 
 chemistry, celebrates its 80-th anniversary in 2005.
Despite that, our theoretical understanding of the PEP origin is still 
not quite satisfactory.  Perhaps the best way to illustrate it is to 
quote Feynman \cite{Feynman}: ``Why is it that particles with 
half-integral spin are Fermi particles whereas particles with integral 
spin are Bose particles? We apologize for the fact that we cannot give 
you an elementary explanation. This probably means that we do not have a 
complete understanding of the fundamental principle involved.'' Similar 
dissatisfactions were expressed by Pauli \cite{Pauli1955} and Dirac 
\cite{Dirac}. 

One important aspect of comprehending a fundamental principle is to 
understand if and how it can be violated and how one can search for its 
violations experimentally. 

No theoretical models of small PEP violation existed up until 1987. 
Although non-standard types of statistics, such as parastatistics 
(\cite{127,Greenberg}), had been known, they could be viewed as ``100\%  
violation'' of PEP rather than small violation. This kind of violation 
for electrons and nucleons  was clearly ruled out by experiment. Also, 
there were studies of possible small violation of electron identity 
(\cite{28,18}; see also \cite{19,27}). Although related to small PEP 
violation, small non-identity is a different idea. 

Altogether, during 30 years prior to 1987 there were about a dozen of 
papers on the topic including pioneering works  \cite{1a,20,2,3} while 
since 1987 the number of papers has grown to several hundreds.  

In 1987 the first quantum-mechanical one-level model of small PEP 
violation (now referred to as Ignatiev-Kuzmin model) was constructed in  
\cite{4}. 

It is based on the trilinear algebra of the creation and annihilation 
operators containing a small parameter  $\beta$:
 \begin{eqnarray}
a^2a^{\dagger} + \beta^2 a^{\dagger}a^2 &=& \beta^2 a\label{eq:9}\\ 
a^2a^{\dagger} + \beta^4a^{\dagger}a^2 &=& 
\beta^2aa^{\dagger}a.\label{eq:10} 
\end{eqnarray}

This model allows for the bilinear particle number operator of the 
following form: 
\begin{equation}
\label{eq:14} N=A_1 a^{\dagger}a + A_2 aa^{\dagger} + A_3  
\end{equation}
with the coefficients $A_i$ given by: 
\begin{eqnarray}
\label{eq:15} A_1 &=& \frac{-1+2\beta^2}{1-\beta^2+\beta^4} \nonumber\\ 
A_2 &=& \frac{-2+\beta^2}{1-\beta^2+\beta^4} \\ A_3 &=& 
\frac{2-\beta^2}{1-\beta^2+\beta^4}. \nonumber 
\end{eqnarray}

Fundamental mathematical properties of the IK algebra have been studied 
in \cite{13,72} (see also \cite{Ignatiev}). These and further studies 
revealed interesting connections with such concepts as Jordan pairs, 
$C^*$-algebras, and quantum groups. 

It is surprising to see parallelism in the development of the two new 
areas: quantum groups in pure mathematics (see, e.g., \cite{qg}) and 
small statistics violation in physics. The fast growth of  both areas  
started independently and about the same time \footnote{The important 
role of quantum groups in modern mathematics has been emphasised by 
awarding the 1990 Fields Medal to V.G.Drinfeld in part for his 
outstanding contributions to that area.}. After he links between them 
were revealed, both fields  benefited from the cross-stimulation. 

Other interesting links between small statistics violation and the rest 
of physics include quantum gravity, anyons, quantum optics, and, more 
recently, non-commutative space-time theories \cite{pinzul}. 

 The problem of a 
realistic generalization of the one-level IK model can be formulated as 
the following question: how to write down the commutation relations if we 
ascribe the momentum and spin indices to the creation and annihilation 
operators? One way to do this was suggested by Greenberg and Mohapatra 
\cite{131,134}. The main problem is to make sure that no more than 2 
electrons can occupy the same state. In the 1-level model that was 
achieved by requiring $a^3=0$. However, in the multi-level model we have 
infinitely many possibilities to put 3 electrons in the same state 
because we can ``sandwich'' other electrons between them. These 
sandwiched states turn out to have negative norms (\cite{7,26,137}). No 
way out of this difficulty has been found and it is believed to be a 
fatal flaw. In principle, the possibility of curing this theory cannot be 
completely ruled out. A well-known example is the ordinary QED which 
involves negative-norm states that are harmless. 

Another generalisation was attempted by Okun \cite{6} who assumed that 
the operators from different levels obey the usual anticommutation 
relations. This model also ran into serious difficulties discussed by the 
author. 

 We would like to stress that the  ``small PEP violation'' is an 
 intuitive concept and one  can try to formalise it in many different ways leading 
 to different theories with their specific experimental predictions. 
 
 In particular, the theory called ``quons''  was proposed in (\cite{138,139,254,100},
  see also \cite{Biedenharn1989,Macfarlane1989}). This theory is based on {\em bilinear}  
 rather than trilinear commutation relations:
 \begin{equation}
 \label{quon}a_k a^{\dagger}_l - q a^{\dagger}_l a_k = \delta_{kl}.
 \end{equation}
 
 The main physical feature of the model is that there is no limit on the number 
 of particle that can occupy the same state, i.e., all types of Young tableaux are
 allowed  for a system of quons. A review of other properties of quons have been
 given in (\cite{Greenberg:2000zy}, see also \cite{Chow}). 
 
 More detailed theoretical and experimental
reviews and further references can be found in \cite{Hilborn, X}.  An 
extensive bibliography with hundreds of references is contained in 
\cite{GillaspyHilborn}.   
 
 Recently it was 
suggested (\cite{Dolgov:2005qi,Dolgov:2005mi}, see also \cite{
Cucurull:1995bx}) that the neutrino statistics may be anomalous (i.e., 
other than Fermi). Some cosmological and astrophysical consequences of 
this hypothesis have been analysed. They include the possibility of 
neutrino forming cold dark matter through Bose condensation  and the 
differences in the Big Bang nucleosynthesis compared to the standard 
scenario. 

These studies can also be considered as an experimental test of the 
spin-statistics connection for neutrinos. We emphasize that the 
motivation to do such tests is especially strong for the case of Majorana 
neutrinos (as opposed to Dirac neutrinos) because in this case there is 
no difference between particle and antiparticle. As a result, only one 
set of creation-annihilation operators is required and therefore the 
difficulty with negative energy disappears. (Of course, in the Dirac case 
two sets of operators are required and the negative energy problem 
persists.) 

Since at the moment none of the existing models of small PEP violation 
can meet all desirable requirements satisfied by the standard model with 
exact PEP, the discussion  in \cite{Dolgov:2005qi,Dolgov:2005mi}   was 
based on the phenomenological approach using the modified kinetic 
equations. As a possible model underlying such an approach, the following 
scheme was suggested \cite{Dolgov:2005qi,Dolgov:2005mi}:  
\be 
\label{1} a_k = c a_k^f + s a_k^b \;\;\;\; c\equiv \cos\gamma, \;\;\;\;s 
\equiv \sin\gamma 
\ee

\begin{subequations}
\be
\label{1a}
\left[a^f_k, (a^f_{k'})^{\dagger}\right]_+ = \delta_{kk'}
\ee

\be
\label{1b} \left[a^b_k, (a^b_{k'})^{\dagger}\right]_- = \delta_{kk'}. 
\ee
\end{subequations}

Let us try to find commutation relations for the operator $a$. For 
simplicity, we discard the momentum and spin index, i.e. consider a 
one-level version of the scheme. Our task is to express the product 
$aa^{\dag}$ in normal form, i.e. in terms of $a^{\dag}a$. Using 
Eq.~(\ref{1}) we can write 
\be
\label{2} 
a a^{\dagger} = c^2 a^f (a^f)^{\dagger} + s^2 a^b (a^b)^{\dagger} + c s \left[a^f (a^b)^{\dagger} + a^b (a^f)^{\dagger}\right]
\ee 

\be
\label{3} a^{\dagger} a = c^2 (a^f)^{\dagger}a^f + s^2  (a^b)^{\dagger} 
a^b + c s \left[ (a^f)^{\dagger} a^b + (a^b)^{\dagger}a^f\right]. 
\ee 
We see that without further assumptions we cannot exclude operators $a^f$ 
and $a^b$ from these equations and write a commutation relation in terms 
of the operators $a$ and $a^{\dag}$ only. Suppose that in addition to 
Eq.~(\ref{1a}) and (\ref{1b}) we postulate the additional relations  
\be
\label{4} 
\left[a^f, (a^b)^{\dagger}\right]_- = 0, \qquad \left[a^b, (a^f)^{\dagger}\right] = 0.
\ee  
Then we can subtract Eq.~(\ref{3}) from Eq.~(\ref{2}) to obtain
\be
\label{5} 
a a^{\dagger} - a^{\dagger} a = c^2 \left[1-2 (a^f)^{\dagger} a^f \right] + s^2 = 1-2 c^2 (a^f)^{\dagger} a^f.
\ee  
Although we succeeded in excluding $a^b$, the $a^f$ is still present. 
This difficulty would persist if, instead of subtracting, we added 
Eq.~(\ref{3}) to Eq.~(\ref{2}). Taking anticommutators instead of 
commutators in Eq.~(\ref{1a}) and Eq.~(\ref{1b}) would not help either. 

Let us try to preserve the symmetry between the operators $a^b$ and $a^f$ 
by introducing the orthogonal combination 
\be
\label{6} d = c a^b - s a^f. 
\ee   
Then,  $a^f$ and $a^b$ can be expressed through $a$ and $d$:

\begin{eqnarray}
\label{7} a^f &=& c a - s d \\ a^b &=& s a + c d.\nonumber 
\end{eqnarray}

 Plugging these into Eq.~(\ref{1a}) and Eq.~(\ref{1b}) we obtain the 
 following relations:
\be
\label{8} 
c^2 a a^{\dagger} - c s (a d^{\dagger} + d a^{\dagger}) + s^2 d d^{\dagger} = 1 - c^2 a^{\dagger} a + c s(a^{\dagger} d + d^{\dagger} a) - s^2 d d^{\dagger}
\ee  
 and
\be
\label{9} s^2 a a^{\dagger} + c s(a d^{\dagger} + d a^{\dagger}) + c^2 d 
d^{\dagger} = 1 + s^2 a^{\dagger} a + c s(a^{\dagger} d + d^{\dagger} a ) 
+ c^2 d^{\dagger} d. 
\ee 
This system of 2 relations is incomplete because we  need to find normal 
forms for 4 operators: $aa^{\dag}$, $ad^{\dag}$, $da^{\dag}$, and 
$dd^{\dag}$. Therefore, 2 more relations are missing. The 2 commutation 
relations Eq.~(\ref{4}) will give us only 1 independent equation, so a 
different assumption is needed. For example, we can impose additional 
commutation relations of the form 
\be
\label{10} a d^{\dagger} = d^{\dagger} a, \qquad d a^{\dagger} = 
a^{\dagger} d. 
\ee    
 These relations close the system of equations for $aa^{\dag}$, 
 $ad^{\dag}$, $da^{\dag}$, and $dd^{\dag}$. The solution is
\be
\label{11} 
(c^2 - s^2) a a^{\dagger} = c^2 - s^2 - (c^4 + s^4) a^{\dagger} a + 2 c^3 s(a^{\dagger} d + d^{\dagger} a) - 2 c^2 s^2 d^{\dagger} d
\ee   
\be
\label{12} (c^2 - s^2) d d^{\dagger} = c^2 - s^2 + 2c^2 s^2 a^{\dagger} a 
- 2 c s^3 (a^{\dagger}d + d^{\dagger} a) + (c^4 + s^4) d^{\dagger} d. 
\ee   
The algebra is now complete, and is an interesting object for further 
study. Because the commutation relations are mixed, it is natural to ask 
first a simpler question. Can one describe a small violation of the Pauli 
principle by means of ``unmixed'' commutation relations of the second 
order? 

One answer to this question is provided by the quon statistics which is 
based on the unmixed bilinear commutation relation, Eq. (\ref{quon}). 
For discussions of the various aspects of quon statistics we refer the 
reader to the existing literature 
\cite{138,139,254,100,Biedenharn1989,Macfarlane1989,Greenberg:2000zy,Chow}. 
For our purposes the most important feature of the quon statistics is 
that it does not lead to any constraints on the representations of the 
permutation group to which quon wave functions belong. That is, the quon 
wave function can transform according to an arbitrary  Young tableau. In 
other words, any number of quons can occupy the same state. 

In that sense, what we have here is ``a small, but radical'' transition 
from the Fermi statistics, in which only {\em one} fermion can occupy a 
state (i.e., only one-column Young tableaux are allowed), to the quon 
statistics where {\em any} number of quons can occupy the same state 
(i.e., any Young tableaux is allowed ). 

It seems natural to ask if there exists a  ``less radical'' non-Fermi 
statistics in which the occupancy of a state $n$ can be greater than 1, 
but still cannot be greater than some integer  $m$:
 \be
\label{14} 0 \leq n \leq m. 
\ee     
The wave functions in such statistics would transform according to the 
Young tables in which the number of columns must not exceed $m$.   
 Let us show that such violation of the 
Pauli principle cannot be described by any algebra defined by {\em 
bilinear} commutation relations.  To be more exact, we shall prove the 
following theorem.\\

{\bf Theorem:} Let the operators $a, a^{\dagger}$ obey a bilinear commutation relation of the form

\begin{equation}
C_1 a^{\dagger}a + C_2 a a^{\dagger} + C_3 1 + C_4 a + C_5 a^{\dagger} + 
C_6 a^2 + C_7 (a^{\dagger})^2 = 0. 
\end{equation}

In addition to it, let there exists an Hermitean opeartor $N$, satisfying 
the commutation relations of the form 

\begin{equation}
 [N,a]=-a, \quad [N,a^{\dagger}]=a^{\dagger} \label{eq:commutators}
\end{equation}

and having its eigenvalues the numbers $0,1,\ldots, n-1$.

Then, for any finite-dimensional representation of the operator $a$, the following equality holds true:

\begin{equation}
  a^2=0.
\end{equation}

{\bf Proof:} Let us work in a representation where the operator $N$ is diagonal

\begin{equation}
  N={\rm diag}(\lambda_1, \ldots, \lambda_n), \quad \lambda_k = k-1,
\end{equation}

$n$ being the dimensionality of the representation.  Let us find the commutator $[N,a]$ in the representation:

\begin{eqnarray}
(Na)_{ij} &=& \sum_k N_{ik} a_{kj} = \lambda_{\underline{i}} a_{\underline{i}j} \nonumber\\
(aN)_{ij} &=& \sum_k a_{ik} N_{kj} = a_{i\underline{j}} \lambda_{\underline{j}} \label{eq:notsummed}\\
  \left[N, a\right]_{ij} &=& a_{\underline{i}\underline{j}}(\lambda_{\underline{i}} - \lambda_{\underline{j}}). \nonumber
\end{eqnarray}

The underlined indices are not summed over.

Substituting Eq.~(\ref{eq:notsummed}) into Eq.~(\ref{eq:commutators}), we obtain the equation

\begin{equation}
  a_{\underline{i}\underline{j}} (\lambda_{\underline{i}} - \lambda_{\underline{j}}) = -a_{ij}. \label{eq:eval}
\end{equation}

From Eq.~(\ref{eq:eval}) it follows that the only nonvanishing elements of matrix $a$ can be $a_{i,i+1}$ ($i=1,\ldots, n-1$), that is, the elements, standing along the over-the-main diagonal (we shall call this diagonal the first one).

Calculating the matrices $a^{\dagger}a$, $a a^{\dagger}$, $a^2$, we find

\begin{eqnarray}
  a^{\dagger}a&=&{\rm diag}(0,|a_{12}|^2,|a_{23}|^2,\ldots, |a_{n-1,n}|^2) \nonumber\\
  a a^{\dagger} &=& {\rm diag}(|a_{12}|^2,|a_{23}|^2,\ldots,0) \label{eq:diag}\\
  (a^2)_{i,i+2}&=&a_{i,i+1} a_{i+1,i+2}, \quad i=1,\ldots,n-2,\nonumber
\end{eqnarray}

the other $(a^2)_{ij}$ being equal to zero.  Thus, all the nonvanishing elements of the matrix $a^2$ stand along the second diagonal (that is the diagonal lying over the first one).

Now consider the bilinear commutation relation of the most general form

\begin{equation}
  C_1 a^{\dagger} a + C_2 a a^{\dagger} + C_3 1 + C_4 a + C_5 a^{\dagger} + C_6 a^2 + C_7  (a^{\dagger})^2 = 0. \label{eq:general}
\end{equation}

All the terms in Eq.~(\ref{eq:general}) are diagonal or quasidiagonal: the terms $C_1-C_3$ are diagonal; the terms $C_4$ and $C_5$ are situated on the first diagonal and its symmetric one; the terms $C_6$ and $C_7$ are situated on the second diagonal and its symmetric one.  From this fact it follows that when $n\geq 3$ the following equalities should hold:

\begin{eqnarray}
  \label{eq:shouldhold}
  C_1 a^{\dagger}a + C_2 a a^{\dagger} + C_3 1 &=& 0, \nonumber\\
 C_4 a=0, \quad C_5 a^{\dagger} &=& 0, \\
 C_6 a^2 =0, \quad C_7 (a^{\dagger})^2 &=& 0.\nonumber
\end{eqnarray}

We are, of course, interested in the case where $a$ and $a^2$ are not equal to zero, therefore the coefficients $C_{4,5,6,7}$ should be equal to zero.  Thus, provided $a, a^2 \neq 0$, the commutation relation in Eq.~(\ref{eq:general}) is equivalent to Eq.~(\ref{eq:shouldhold}).

Substituting into Eq.~(\ref{eq:shouldhold}) the explicit forms of the operators $a^{\dagger} a$ and $a a^{\dagger}$, we obtain the set of $n$ equations for the determination of the coefficients $C_1-C_3$ and matrix elements $a_{i,i+1}$:

\begin{eqnarray}
  \label{eq:coefficients}
  C_2 b_1 + C_3 &=& 0 \nonumber\\
  C_1 b_1 + C_2 b_2 + C_3 &=& 0\nonumber\\
  C_1 b_2 + C_2 b_3 + C_3 &=& 0\\
  C_1b_{n-1} + C_3 &=& 0,\nonumber
\end{eqnarray}

where $b_i=a_{i,i+1}^2, b_n=0$.

This set of equations can be considered as the set of $n$ linear equations with three unknowns $C_1-C_3$, the coefficients $b_i$ being non-negative and containing at least one pair of neighbouring nonzero numbers: $b_kb_{k+1}\neq 0$ (the latter condition is equivalent to the requirement $a^2\neq0$).  In order that this set has a nonzero solution (i.e. not all $C_i$ were vanishing) it is necessary that the rank of the coefficient matrix $M_n$ was less than three (i.e. was equal to 1 or 2), where

\begin{equation}
  \label{eq:Mn}
  M_n = \left| \begin{array}{ccc}
0 & b_1 & 1 \\
b_1 & b_2 & 1 \\
b_2 & b_3 & 1 \\
\vdots & \vdots & \vdots \\
b_{n-1} & b_n & 1 
\end{array} \right|
\end{equation}

Let us make two transformations of the matrix $M_n$ which will not change its rank, namely: subtract the first line from all the other lines and then subtract the third column multiplied by $b_1$ from the second column. Then the matrix will take the form

\begin{equation}
  \label{eq:Mndash}
  M'_n = \left| \begin{array}{ccc}
0 & 0 & 1 \\
b_1 & b_2-b_1 & 0 \\
b_2 & b_3-b_1 & 0 \\
\vdots & \vdots & \vdots \\
b_{n-1} & b_n-b_1 & 0
\end{array} \right|
\end{equation}

The rank of this matrix is evidently larger by one than the rank of the matrix

\begin{equation}
  \label{eq:mndashdash}
    M''_n = \left| \begin{array}{cc}
b_1 & b_2-b_1  \\
b_2 & b_3-b_1  \\
\vdots & \vdots  \\
b_{n-1} & b_n-b_1
\end{array} \right|
\end{equation}

For the following considerations it is convenient to exclude the first three special cases: $b_1=0; b_1\neq0; b_2/b_1=1; b_1\neq 0; b_2/b_1=2$.

{\em The case $b_1 = 0$}. Then the matrix $M''_n$ takes the form

\begin{equation}
  \label{eq:mndashdashcaseOne}
    M''_n = \left| \begin{array}{cc}
0 & b_2  \\
b_2 & b_3  \\
\vdots & \vdots  \\
b_{n-1} & b_n
\end{array} \right|
\end{equation}

We require the linear dependence of the second and first lines and thus we obtain $b_2=0$; then, analogously, $b_3=0$ etc., and, finally, $b_{n-1}=0$ which means that all $b_i$ are vanishing.  Therefore, the case $b_1=0$ is excluded.

{\em The case $b_1\neq 0, b_2/b_1=1$.}  Acting in analogy with the previous case, one can prove, that the linear dependence of all the lines implies that all $b_k$ (including $b_n$) are equal to $b_1$.  This result, however, contradicts the condition $b_n=0$.

{\em The case $b_1\neq0, b_2/b_1=2$.}  In this case for all $b_k$ (including $b_n$) it is true that $b_k=kb_1$ which again contradicts the condition $b_n=0$.

Having excluded the above special cases, consider now the most general case, when $b_1\neq 0$ and $b_1/b_2\neq 1,2$.  Write down the condition of linear dependence of the $k$-th and the first lines of the matrix $M''_n$ (using the condition $b_1\neq 0$).

\begin{eqnarray}
  \label{eq:detImplies}
  \det \left|\begin{array}{cc}
b_1 & b_2-b_1 \\
b_k & b_{k+1} - b_1 
\end{array} \right| &=& 0 \\
 \Rightarrow b_{k+1} - b_k \left(\frac{b_2-b_1}{b_1}\right) - b_1 &=& 0.\nonumber
\end{eqnarray}

In order to eliminate the inhomogeneity in this recurrent equation, let us shift $k\rightarrow k+1$:

\begin{equation}
  b_{k+2} - b_{k+1} \left(\frac{b_2-b_1}{b_1}\right) - b_1 = 0.\label{eq:postShift}
\end{equation}

and, subtracting Eq.~(\ref{eq:detImplies}) from Eq.~(\ref{eq:postShift}), obtain

\begin{equation}
  \label{eq:subtracted}
  b_{k+2} - b_{k+1} \frac{b_2}{b_1} + b_k \left( \frac{b_2-b_1}{b_1}\right) = 
  0.
\end{equation}

The characteristic equation for this recurrent relation has the form 

\begin{equation}
 x^2 - \frac{b_2}{b_1} x + \frac{b_2-b_1}{b_1}=0.
\end{equation}

Its solutions are 

\begin{equation}
x_1 = \frac{b_2-b_1}{b_1}, \quad x_2 = 1.
\end{equation}

Therefore, the general solution of Eq.~(\ref{eq:subtracted}) can be written in the form

\begin{equation}
  \label{eq:generalSoln}
b_k = C_1 x_1^k + C_2 x_2^k = C_1 \left( \frac{b_2-b_1}{b_1}\right)^k + 
C_2.  
\end{equation}

The constants $C_1$ and $C_2$ can be determined from the initial conditions: the first two terms of the sequence Eq.~(\ref{eq:generalSoln}) are equal to $b_1$ and $b_2$, hence, using the conditions $b_1/b_2=1,2$ we find

\begin{eqnarray}
 C_1 &=& \frac{b_1b_2}{b_1-2b_2} \\
 C_2 &=& \frac{-b_1b_2}{b_1-2b_2}.
\end{eqnarray}

So, the $n$-th term of the sequence Eq.~(\ref{eq:generalSoln}), which we are interested in, has the form

\
\begin{equation}
  b_n = \frac{b_1b_2}{b_1-2b_2} \left[ 
  \left(\frac{b_2}{b_1}-1\right)^n-1\right].
\end{equation}

Equating $b_1$ to zero, we find that (taking into account the condition $b_1/b_2=2$) for odd $n$ the equation $b_n=0$ has no solution whereas for even $n$ there is the unique solution $b_2/b_1=0$, therefore

\begin{equation}
  b_{\rm even} = 0, \quad b_{\rm odd} = b_1.
\end{equation}

This sequence, however, does not contain a pair of neighbouring nonzero 
terms.  Thus the last case under consideration is also excluded. Q.E.D. 

Thus we have proved a no-go theorem which forbids to have a   finite 
occupancy limit for an arbitrary system with a bilinear commutation 
relation. In other words, either the upper limit on the occupancy number 
is 1 (the ordinary fermionic case) or there is no upper limit at all. 
Some examples of the latter class include the usual Bose statistics, as 
well as non-standard quon statistics and infinite statistics.    

  We are 
grateful to A.Yu.Smirnov and B.H.J.McKellar for  interesting discussions. 
This work was supported in part by the Australian Research Council.

\end{document}